# Implications of Atmospheric Nondetections for Trappist-1 Inner Planets on Atmospheric Retention Prospects for Outer Planets

Joshua Krissansen-Totton
Department of Earth and Space Sciences/Astrobiology Program, University of Washington, Seattle, WA, USA; joshkt@uw.edu

## Abstract

JWST secondary eclipse observations of Trappist-1b seemingly disfavor atmospheres $>\sim$1 bar since heat redistribution is expected to yield dayside emission temperature below the $\sim$500 K observed. Given the similar densities of Trappist-1 planets, and the theoretical potential for atmospheric erosion around late M dwarfs, this observation might be assumed to imply substantial atmospheres are also unlikely for the outer planets. However, the processes governing atmosphere erosion and replenishment are fundamentally different for inner and outer planets. Here, an atmosphere–interior evolution model is used to show that an airless Trappist-1b (and c) only weakly constrains stellar evolution, and that the odds of outer planets e and f retaining substantial atmospheres remain largely unchanged. This is true even if the initial volatile inventories of planets in the Trappist-1 system are highly correlated. The reason for this result is that b and c sit unambiguously interior to the runaway greenhouse limit, and so have potentially experienced $\sim$8 Gyr of X-ray and extreme ultraviolet–driven hydrodynamic escape; complete atmospheric erosion in this environment only weakly constrains stellar evolution and escape parameterizations. In contrast, e and f reside within the habitable zone, and likely experienced a comparatively short steam atmosphere during Trappist-1's pre-main sequence, and consequently complete atmospheric erosion remains unlikely across a broad swath of parameter space (e and f retain atmospheres in $\sim$98% of model runs). Naturally, it is still possible that all Trappist-1 planets formed volatile-poor and are all airless today. But the airlessness of b (and c) does not require this, and as such, JWST transit spectroscopy of e and f remains the best near-term opportunity to characterize the atmospheres of habitable zone terrestrial planets.

*Unified Astronomy Thesaurus concepts:* Exoplanets (498); Habitable planets (695); Exoplanet astronomy (486)

## 1. Introduction

Whether the Trappist-1 planets and other terrestrial planets around late M dwarfs typically possess atmospheres is unknown. The theoretical challenges to atmospheric retention are numerous. Late M dwarfs undergo an extended, superluminous pre-main-sequence phase, during which time high X-ray and extreme ultraviolet (XUV) and bolometric fluxes can drive high thermal escape rates (Luger & Barnes 2015; Bolmont et al. 2017; Bourrier et al. 2017; Wordsworth et al. 2018; Barth et al. 2021). Moreover, since small orbital separations are implied for "Venus Zone" (Ostberg & Kane 2019) and habitable zone (Kopparapu et al. 2013) instellations, terrestrial planets may also experience high nonthermal loss rates due to the stellar wind (Garcia-Sage et al. 2017; Dong et al. 2018). Atmospheric erosion could also be enhanced by more frequent flaring and high stellar activity (do Amaral et al. 2022). The high velocity impacts for planets around M dwarfs could result in severe impact erosion of atmospheres (Kral et al. 2018), although excessive impact erosion is seemingly unlikely in the Trappist-1 system due to fragility of the observed orbital resonances, which places constraints on cumulative impactor fluxes (Raymond et al. 2022). While the possibility for complete atmospheric erosion exists, initial volatile endowments are potentially large (Ormel et al. 2017; Coleman et al. 2019; Miguel et al. 2020), and volatile replenishment from the deep interior via magmatic degassing may be substantial. Moreover, the low bulk densities of the Trappist-1 planets, particularly f and g, seemingly require either a large surface volatile inventory and/or an anomalously small or low density metallic core relative to solar system objects (Agol et al. 2021; Raymond et al. 2022; Schlichting & Young 2022). When compared to a broader population of planetary objects, the Trappist-1 planets (and other late M dwarf hosted terrestrials) reside approximately on the so-called "cosmic shoreline" that empirically separates airless worlds and planets with substantial atmospheres (Zahnle & Catling 2017). The characterization of objects on the cosmic shoreline may help constrain atmospheric loss processes more generally.

Terrestrial planet observations in the pre-JWST era provided considerable insight into atmospheric erosion. In particular, thermal emission observations of the highly irradiated rocky planets LHS 3844 b and GJ 1252 b revealed dayside temperatures consistent with the lack of atmospheric heat redistribution, which has been interpreted to imply a lack of any substantial atmosphere (Kreidberg et al. 2019; Crossfield et al. 2022). These examples of atmospheric loss can be contrasted with tentative atmospheric detections on the ultrahot rocky planet 55 Cancri e via thermal phase curve observations (Demory et al. 2016; Angelo & Hu 2017; Hammond & Pierrehumbert 2017). With a dayside temperature of $\sim$2700 K, the presence of an atmosphere on 55 Cancri e would suggest bolometric instellation is not the only determinant of atmospheric retention. The apparent lack of $H_2$-atmospheres on the Trappist-1 planets (De Wit et al. 2018; Turbet et al. 2020) and GJ 1132b (Diamond-Lowe et al. 2018; Mugnai et al. 2021; Libby-Roberts et al. 2022) broadly confirm theoretical expectations that $H_2$ atmospheres on Earth-sized planets are susceptible to XUV-driven loss (Owen et al. 2020).

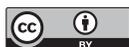

JWST provides improved precision and wavelength coverage for transit spectroscopy and thermal emission observations, and a corresponding potential to transform our understanding of atmospheric origins, loss, and replenishment on rocky exoplanets. For planets that have retained substantial atmospheres, JWST may be capable of molecular detections and abundance constraints of $CO_2$, $H_2O$, $CH_4$, CO, and $O_2$ (Krissansen-Totton et al. 2018b; Lustig-Yaeger et al. 2019; Wunderlich et al. 2019, 2020; Gialluca et al. 2021). Indeed, JWST transit spectroscopy observations of the newly discovered LHS 475b confirm these anticipated capabilities (Lustig-Yaeger et al. 2023). Simultaneously, techniques for rapidly triaging terrestrial exoplanets for the presence or absence of an atmosphere using JWST secondary eclipse observations are in use (Koll et al. 2019), and for planets that have lost their atmospheres, characterizing surface mineralogy in thermal emission may be feasible (Hu et al. 2012; Whittaker et al. 2022).

Thermal emission observations of Trappist-1b were recently used to infer a dayside brightness temperature of $503^{+26}_{-27}$ K (Greene et al. 2023). This is consistent with reradiation from an airless dayside as substantial atmospheres (>1 bar) would redistribute heat resulting in a lower dayside brightness temperature (Greene et al. 2023). Follow-up work (Ih et al. 2023) modeled the dayside thermal emission of Trappist-1b for a wide range of atmospheric compositions and confirmed that >1 bar atmospheres containing even trace amounts of $CO_2$ are strongly disfavored. Dense atmospheres (>1 bar) consisting of predominantly nonabsorbing or weakly absorbing species ($N_2$, $O_2$, CO) can potentially be reconciled with the observed brightness temperature, but such atmospheres are geochemically implausible given that complete photochemical conversion of $CO_2$ to CO is unlikely (Gao et al. 2015) and that even highly reduced silicate melts will degas some carbon as $CO_2$ (Sossi et al. 2020). Upcoming phase curve observations of Trappist-1b (Cycle 2 GO Proposal 3077) may confirm or refute the absence of a substantial atmosphere.

The high dayside brightness temperature of Trappist-1b, taken in combination with evidence for lack of atmospheres around other hot, M dwarf hosted terrestrial planets (Kreidberg et al. 2019; Crossfield et al. 2022), could be interpreted to disfavor atmospheres on the outer Trappist-1 planets. Previous applications of a coupled atmosphere-evolution model (Krissansen-Totton & Fortney 2022) predicted that inner planets Trappist-1b and c lost their atmospheres after ~8 Gyr of evolution in about half of all simulations, whereas the outer planets Trappist-1e and f almost always retained substantial atmospheres. Here, I update these calculations for e and f given recent JWST constraints on the lack of an atmosphere on b. I argue that the airlessness of b (and c) does not substantially modify the odds of e and f also being airless.

## 2. Methods

A previously published version of the Planetary Atmospheres, Crust, MANtle (PACMAN) geochemical evolution model was adopted to simulate the evolution of the Trappist-1 planets (Krissansen-Totton & Fortney 2022). PACMAN captures the key processes that shape atmospheric evolution from magma ocean origins to potentially more temperate states (Krissansen-Totton et al. 2021b; Krissansen-Totton & Fortney 2022). The model explicitly couples the time evolution of mantle oxidation state, atmosphere–interior volatile exchange—including degassing, surface weathering, and oxidation reactions—parameterized mantle convection, and thermal and nonthermal atmospheric escape to compute how atmospheric composition and surface climate evolve on Gyr timescales. PACMAN has previously been applied to successfully model the evolution of Earth (Krissansen-Totton et al. 2021b), Venus (Krissansen-Totton et al. 2021a), and to make preliminary predictions for the Trappist-1 planets (Krissansen-Totton & Fortney 2022). Crucially, PACMAN can assess the statistical likelihood of different atmospheric outcomes via Monte Carlo calculations that sample unknown initial compositions and uncertain model parameters (e.g., Table 1). Figure 1 shows a schematic of PACMAN highlighting how volatile exchange, energy budgets, and redox processes are incorporated during both magma oceans and temperate evolution phases.

The PACMAN model remains unchanged from Krissansen-Totton & Fortney (2022; hereafter KTF22) except that I resample input parameters to reflect recent JWST observations. Specifically, both Trappist-1b and c evolution calculations are run simultaneously, and selected parameters describing stellar evolution, atmospheric escape, and atmosphere–surface interactions are assumed to apply to both planets (Table 1). Of those Monte Carlo calculations for b and c, we focus on only the model runs that result in both b and c being airless, defined as <1 bar total surface volatile inventories of $CO_2$, $H_2O$, and $O_2$ (see Discussion for an explanation of this definition). The sets of common parameter values that produce an airless b and c are subsequently resampled, and applied to compute the atmospheric evolution of e and f. In other words, the stellar evolution, atmospheric escape, and atmosphere–surface interaction parameters values that ensure an airless b and c are applied to e and f to test the extent to which they constrain the atmospheric evolution of these worlds. This approach assumes the initial volatile inventories of the Trappist-1 planets are completely independent. To test the other endmember case, we repeat the calculations described for e and f, except that initial volatile inventories for $H_2O$, $CO_2$, and free O are resampled from the inferred initial volatile distributions for b that ensure an airless outcome. This effectively assumes that initial volatile inventories for the outer planets are drawn from the same distribution as initial volatile endowments for the inner planets that produce airlessness.

## 3. Results

Figure 2 shows the distribution of modern surface pressures for Trappist-1e and f in the scenarios described above. First, I show the predictions from KTF22, which were made before any JWST observations of the Trappist-1 system (denoted "Prior"). Alongside this distribution we show the surface pressure outcomes when only common stellar, escape, and atmosphere–surface interaction parameters that recover an airless Trappist-1b and c are resampled for the outer planets (denoted "Airless b&c"). The likelihood of complete atmospheric erosion, as defined by the fraction of model runs with final surface volatile inventories <1 bar, remains approximately unchanged for e (0.9% to 0.8%), and increases slightly for f (0.7% to 1.4%), but overall atmospheric retention is favored. Similarly, when the conditions producing an airless b and c are imposed, and the initial volatile inventories for e and f are

**Table 1**
Uncertain Parameter Ranges Sampled in Nominal Trappist-1 Monte Carlo Calculations

| | | Nominal Range | References/Notes |
|---|---|---|---|
| Initial conditions | Water[a] | $10^{21}$–$10^{23.63}$ kg[b] | 0.7–300 Earth oceans, or 0.02–7 wt% water for an Earth-mass planet |
| | Carbon dioxide[a] | $10^{20}$–$10^{22.69}$ kg[b] | Approximately 20 bar–10 kbar, pending other atmospheric constituents and gravity. |
| | Radionuclide U, Th, and K inventory (relative to Earth) | 0.33–30.0[b] | Scalar multiplication of Earth's radionuclide inventories in Lebrun et al. (2013). Allows for modest tidal heating. |
| | Mantle free oxygen[a] | $10^{20.6}$–$10^{22}$ kg[b] | This ensures postsolidification mantle redox within 3–4 log units of the quartz–fayalite–magnetite buffer. |
| Stellar evolution and escape parameters | **Trappist-1 XUV saturation time, $t_{sat}$** | $3.14^{+2.22}_{-1.46}$ Gyr | **XUV evolution parameters drawn randomly from joint distribution (Birky et al. 2021).** |
| | **Post saturation phase XUV decay exponent, $\beta_{decay}$** | $-1.17^{+0.27}_{-0.28}$ | **XUV evolution parameters drawn randomly from joint distribution (Birky et al. 2021).** |
| | **Saturated $\log_{10}(F_{XUV}/F_{BOLOMETRIC})$ flux ratio** | $-3.03^{+0.25}_{-0.23}$ | **XUV evolution parameters drawn randomly from joint distribution (Birky et al. 2021).** |
| | Escape efficiency at low XUV flux, $\varepsilon_{low}$ | 0.01–0.3 | See escape section in Krissansen-Totton et al. (2021b). |
| | Transition parameter for diffusion limited to XUV-limited escape, $\lambda_{tra}$ | $10^{-2}$–$10^{2}$[b] | See escape section in Krissansen-Totton et al. (2021b). |
| | XUV energy that contributes to XUV escape above hydrodynamic threshold, $\zeta_{high}$ | 0%–100% | See escape section in Krissansen-Totton et al. (2021b). |
| | Cold trap temperature variation, $\Delta T_{cold-trap}$ | $-30$ to $+30$ K | Cold trap temperature, $T_{cold-trap}$, equals planetary skin temperature plus a fixed, uniformly sampled variation, $T_{cold-trap} = T_{eq}(1/2)^{0.25} + \Delta T_{cold-trap}$. Here, $T_{eq}$ is the planetary equilibrium temperature given assumed albedo. |
| | Thermosphere temperature, $T_{thermo}$ | 200–5000 K[b] | (Lichtenegger et al. 2016; Johnstone et al. 2018, 2021) |
| | Nonthermal escape (total loss over Trappist-1 evolution), $NT$ | 1–100 bar[b] | (Garcia-Sage et al. 2017; Dong et al. 2018) |
| Carbon cycle parameters | Temperature-dependence of continental weathering, $T_{efold}$ | 5–30 K | Plausible Earth-like range (Krissansen-Totton et al. 2018a) |
| | $CO_2$-dependence of continental weathering, $\gamma$ | 0.1–0.5 | Plausible Earth-like range (Krissansen-Totton et al. 2018a) |
| | Weathering supply limit, $W_{sup-lim}$ | $10^{5}$–$10^{7}$ kg s$^{-1}$[b] | Broad terrestrial planet range (Foley 2015) |
| | Ocean calcium concentration, $[Ca^{2+}]$ | $10^{-4}$–$3 \times 10^{-1}$ mol kg$^{-1}$[b] | Plausible range for diverse terrestrial planet compositions (Kite & Ford 2018; Krissansen-Totton et al. 2018a) |
| | Ocean carbonate saturation, $\Omega$ | 1–10 | (Zeebe & Westbroek 2003) |
| Interior evolution parameter | Solid mantle viscosity coefficient, $V_{coef}$ | $10^{1}$–$10^{3}$ Pa s | Solid mantle kinematic viscosity, $\nu_{rock}$, (m$^2$ s$^{-1}$) is given by the following equation: $\nu_{rock} = V_{coef} 3.8 \times 10^7 \exp\left(\frac{350000}{8.314 T_p}\right)/\rho_m$ Here $T_p$ is mantle potential temperature (K) and $\rho_m$ is mantle density (kg m$^{-3}$). See Krissansen-Totton et al. (2021b). |
| Crustal sinks oxygen and hydrological cycle parameters | Crustal hydration efficiency, $fr_{hydr-frac}$ | $10^{-3}$ to 0.03[b] | Upper limit wt % $H_2O$ in oceanic crust. Lower limit hydration limited by cracking. |
| | **Dry oxidation efficiency, $f_{dry-oxid}$** | **$10^{-4}$ to 10%[b]** | **Plausible range of processes for Venus (Gillmann et al. 2009)** |
| | Wet oxidation efficiency, $f_{wet-oxid}$ | $10^{-3}$–$10^{-1}$[b] | Based on oxidation of Earth's oceanic crust (Lécuyer & Ricard 1999). |
| | **Maximum fractional molten area, $f_{lava}$** | **$10^{-4}$–1.0[b]** | **See explanation in Krissansen-Totton et al. (2021b).** |
| | Max mantle water content, $M_{solid-H_2O-max}$ | 0.5–15 Earth oceans | Best estimates maximum hydration of silicate mantle (Cowan & Abbot 2014) |
| | | 0.0–0.2 | (Pluriel et al. 2019) |

**Table 1**
(Continued)

|  |  | Nominal Range | References/Notes |
| --- | --- | --- | --- |
| Albedo parameters | Hot state albedo (during runaway greenhouse/magma ocean), $A_H$ |  |  |
|  | Cold state albedo (during temperate state), $A_C$ | 0.0–0.5 | (Shields et al. 2013; Kopparapu et al. 2017; Rushby et al. 2020; Macdonald et al. 2022) |

**Notes.** Bold parameters are assumed to be common to all planets in the Trappist-1 system, whereas other parameters are sampled independently for each planet in the system.
[a] Denotes variables sampled independently for all planets in nominal calculations, but drawn from the distribution that ensures an airless Trappist-1b subsequently.
[b] Denotes this variable was sampled uniformly in log space. All others (except stellar XUV parameters) sampled uniformly in linear space.

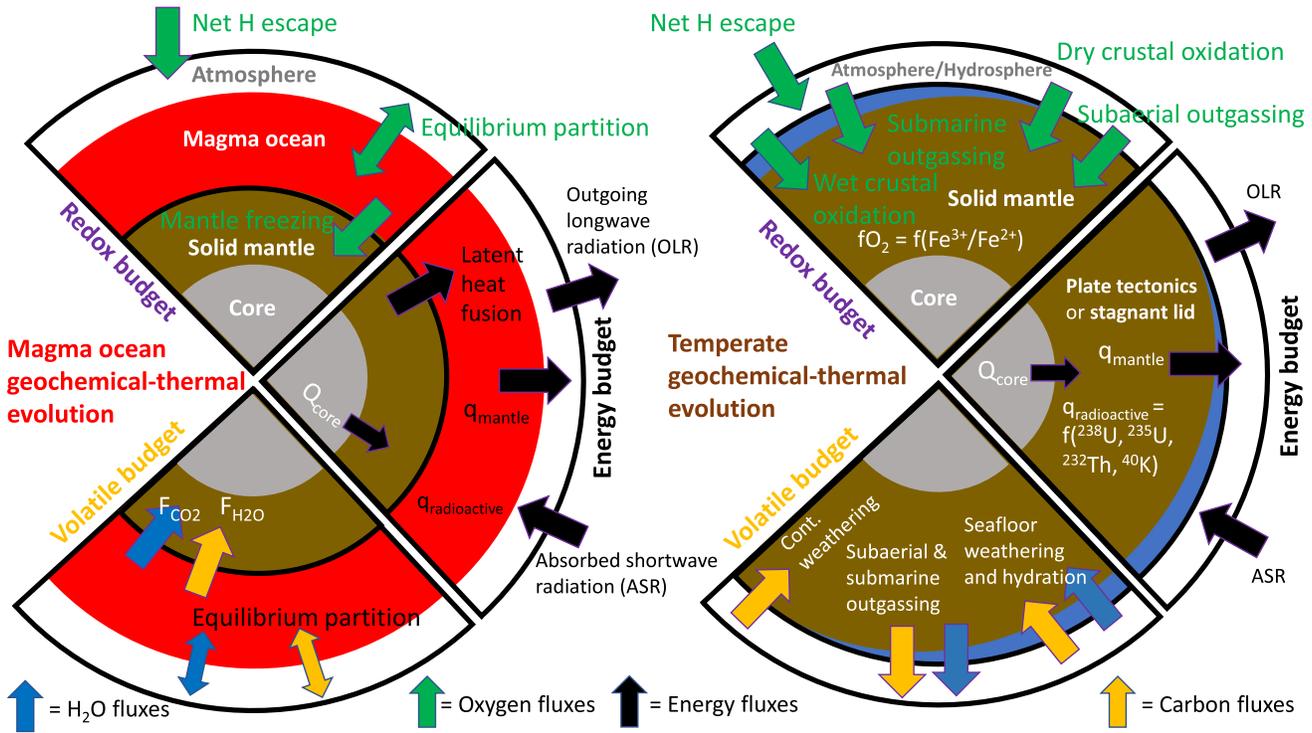

**Figure 1.** Schematic of the PACMAN geochemical evolution model. The redox budget, thermal-climate evolution, and volatile budget are modeled from postaccretion magma ocean (left) through to temperate geochemical cycling (right). Oxygen fluxes are shown by green arrows, energy fluxes by black arrows, carbon fluxes by orange arrows, and water fluxes by blue arrows; the loss of hydrogen to space effectively adds oxygen to the atmosphere. During the magma ocean phase, the radius of solidification begins at the core–mantle boundary and moves toward the surface as internal heat is dissipated. The rate at which this occurs is controlled by radiogenic and tidal heat production, convective heat flow from the mantle to the surface, and heat flow from the core. During the temperate phase, outgassing, weathering, and oxidation reactions control the exchange of volatiles between the atmosphere and the interior. Internal heat flow balances the difference between outgoing longwave radiation (OLR) and absorbed stellar radiation (ASR) throughout.

resampled from the initial volatile inventories of b and c that result in airless outcomes, atmospheric retention is still strongly favored (denoted "Airless b&c and correlated init. volatiles" in Figure 2). Here, the likelihood of an airless outcome for e and f increases to 2.1% and 2.2%, respectively.

The reason for this behavior is illustrated in Figure 3, which shows the time evolution of all model runs for b that result in an airless planet after ∼8 Gyr. Since b orbits well interior to the habitable zone of Trappist-1, the received bolometric insellation (currently four times Earth's insolation) exceeds the runaway greenhouse threshold throughout its evolution (Figure 3(d)). The resulting steam-rich atmosphere can maintain high hydrodynamic H escape fluxes throughout the lifetime of Trappist-1b (and large associated O and $CO_2$ drag fluxes). Large initial volatile inventories can be completely eroded, as shown in Figure 4 which compares the initial $H_2O$ and $CO_2$ volatile endowments for b that were assumed previously in KTF22, and the initial volatile distributions that satisfy the constraint that b has a negligible modern atmosphere. We observe that only extremely large (>100 Earth oceans) endowments of water are ruled out because there is the potential for large cumulative H loss across 8 Gyr of runaway greenhouse in a high-XUV environment. Initial $CO_2$ endowments are constrained more tightly, but even so, when the constrained distributions in Figures 4(a) and (b) are resampled for Trappist-1e, the result is that an atmosphere is retained in most cases, as described above (Figure 2). This outcome is explained by Figure 5, which shows all model outputs for 1e given the common stellar, escape, and atmosphere–interior interaction parameters that yield an airless b and c, as well an assumed initial volatile distribution identical to that of Trappist-1b (i.e., that of Figure 4). Since e resides unambiguously beyond the runaway greenhouse limit after a few 100 Myr the potential for continuous H loss (and hydrodynamic drag of heavier species like O and $CO_2$) is greatly diminished. Moreover, once surface water has condensed, $CO_2$ may be rapidly weathered out of the atmosphere are sequestered in the interior, where it is shielded from escape and can be subsequently outgassed (Figure 5). Consequently, substantial surface volatiles are retained in all but a handful of model runs.

## 4. Discussion

Taken at face value, the calculations above suggest there is a strong possibility that Trappist-1e and f have retained substantial atmospheres. Atmospheric detections for Trappist-1e and f are potentially more challenging than for b and c since thermal emission techniques require longer integration times at cooler equilibrium temperatures (Koll et al. 2019; Lustig-Yaeger et al. 2019). While molecular detections of $CH_4$, $H_2O$, $CO_2$, or $O_2$ via transmission spectroscopy are feasible for Trappist-1e, and could potentially detect atmospheres in a handful of transits (Krissansen-Totton et al. 2018b; Lustig-Yaeger et al. 2019; Wunderlich et al. 2019, 2020; Gialluca et al. 2021), nonocculted star spots can potentially confound unambiguous detections (Rackham et al. 2018; Rackham & de Wit 2023). High-altitude aerosols may also preclude atmospheric detections via transit spectroscopy (Fauchez et al. 2019; Gao et al. 2021), although photochemical hazes are seemingly less likely on temperate terrestrial worlds (Yu et al. 2021), especially in oxidizing atmospheres (Krissansen-Totton et al. 2016). Regardless, Trappist-1e and f remain promising candidates for atmospheric characterization. Similar arguments apply to g and h, which are also promising candidates for atmospheric characterization. However, for these planets it is alternatively plausible that they retain a large surface volatile inventory but nonetheless possess a nearly airless surface due to the condensation of volatiles such as $H_2O$ and $CO_2$ (Turbet et al. 2018, 2020), and the subsequent nonthermal erosion of noncondensibles like $N_2$ and CO, a possibility we do not model here.

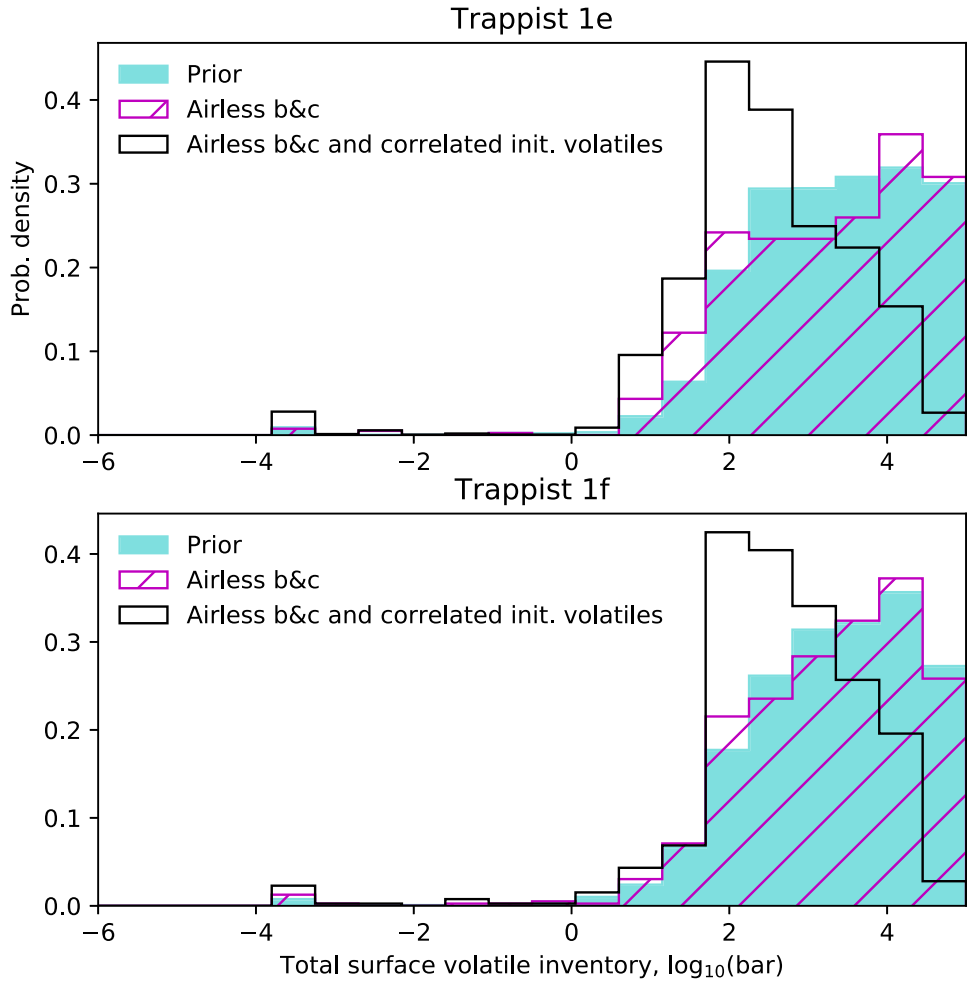

**Figure 2.** JWST observations of an airless Trappist-1b do not substantially change the odds of e and f being airless. Probability distribution of the total surface volatile inventories of Trappist-1e and f. Cyan distributions represent previous PACMAN predictions from KTF22. Magenta distributions show updated results applying the constraints on stellar evolution, atmospheric escape, and atmosphere–surface interactions implied by an airless Trappist-1b; here, initial volatile inventories for all Trappist-1 planets are assumed to be independent. Black distributions show the same updated results, but the initial volatile inventories of e and f are sampled from the same initial volatile distributions that ensure an airless b and c. For both these endmember cases, the distribution of final surface pressures is minimally modified from the original predictions.

Indeed, one important caveat to the calculations presented above is that PACMAN only explicitly considers the erosion of $H_2O$, $CO_2$, and $O_2$, because these are likely the most abundant volatiles based on cosmochemical abundances and volatile partitioning during accretion. For all evolutionary calculations, background atmospheric $N_2$ is fixed at 1 bar, and I assume planets became susceptible to complete atmosphere erosion when total surface inventories of other volatiles (sum of $H_2O$, $CO_2$, and $O_2$) drop below 1 bar. Sensitivity tests presented in KTF22 show that evolutionary calculations are largely insensitive to the assumed background $N_2$ level, within factors of a few. However, it is possible that temperate planets could experience rapid water loss if atmospheric $CO_2$ is sequestered in carbonates and the $N_2$ background is eroded by nonthermal processes; the resulting $H_2O$-dominated atmosphere will have a weak cold trap, potentially leading to high H escape rates (Wordsworth & Pierrehumbert 2014; Kleinböhl et al. 2018). Future versions of PACMAN will explicitly incorporate $N_2$ cycling and N escape, potentially allowing a more complete assessment of the likelihood of atmospheric erosion.

The airlessness of b and c would imply that d is somewhat more likely to be airless since d straddles the inner edge of the habitable zone. Specifically, the percentage of airless outcomes for d increases from 25% to 30% in KTF22 to 44% when "airless b & c" stellar constraints are applied, and 50% when the same initial volatile inventories are sampled (not shown).

Finally, note that the parameterized treatment of atmospheric escape in PACMAN necessarily makes simplifying assumptions. For example, thermosphere temperature, which is randomly sampled from 200 to 5000 K, is assumed to be independent of atmospheric composition and independent for each planet, whereas nonthermal escape fluxes are merely sampled from plausible ranges (Garcia-Sage et al. 2017; Dong et al. 2018). It is possible that more complete escape parameterizations would modify our results, although I emphasize that thermosphere temperature and nonthermal escape rates are not correlated with atmospheric retention for the outer planets (KTF22).

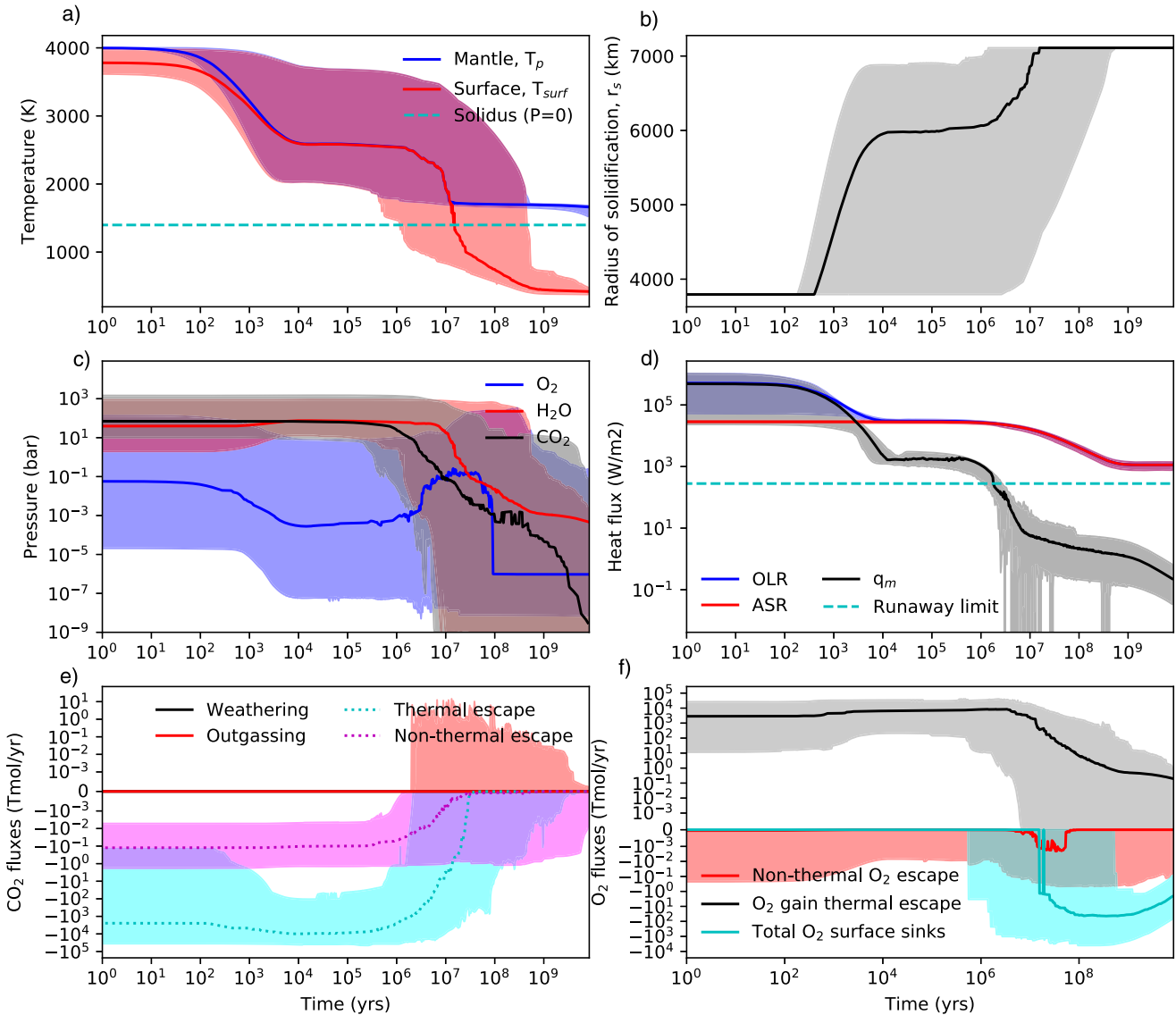

**Figure 3.** Complete atmosphere erosion on Trappist-1b is unsurprising given the potential for continuous hydrodynamic escape until virtually all water is lost. All Trappist-1b evolutionary model outputs consistent with the absence of substantial surface volatiles today (<1 bar). Subplots denote (a) the evolution of mantle (blue) and surface temperature (red), (b) the solidification radius as it moves from the core–mantle boundary to the surface, (c) atmospheric abundances of water (red), oxygen (blue), and carbon dioxide (black), (d) the planetary energy budget including absorbed shortwave radiation (ASR; red), outgoing longwave radiation (OLR; blue), and interior heat flow (black), (e) carbon dioxide fluxes including magmatic outgassing (red), thermal escape (cyan), nonthermal escape (magenta), and weathering (black), and (f) oxygen fluxes including net oxygen gain via hydrogen escape (black), nonthermal $O_2$ loss (red), and $O_2$ consumption by crustal sinks (cyan). Shaded regions denote 95% confidence intervals across all model outputs consistent with the hypothetical lack of an atmosphere.

*What would an airless e or f imply?* Of course, the initial volatile inventories sampled in all our evolutionary calculations are necessarily arbitrary, and it remains possible that all Trappist-1 planets are airless because they formed comparatively volatile depleted (Coleman et al. 2019; Raymond et al. 2022), or because the bulk of their accreted volatiles were sequestered in a metallic core (Schlichting & Young 2022). Figure 6 shows the distribution of initial volatile inventories for e and f that result in total surface volatile depletion after 8 Gyr and compares this to the assumed distribution (including both a "Prior" assumed range and the "Posterior" range of initial inventories consistent with an airless b and c). As expected, an airless b and c does not constrain the initial volatile inventory of e or f due to the possibility of large condensed surface water inventories and the weak implied constraints on stellar evolution. However, if future JWST observations were to confirm that e or f did not possess a substantial surface volatile inventory, then this would put stringent upper limits on their initial water and $CO_2$ endowments (Figure 6), as well potentially on atmospheric escape and atmosphere–interior exchange.

## 5. Conclusions

The airlessness of Trappist-1b only weakly constrains Trappist-1 stellar evolution, atmospheric escape, and atmosphere–surface volatile exchange on hot terrestrial planets. Consequently, when these constraints are applied to the outer

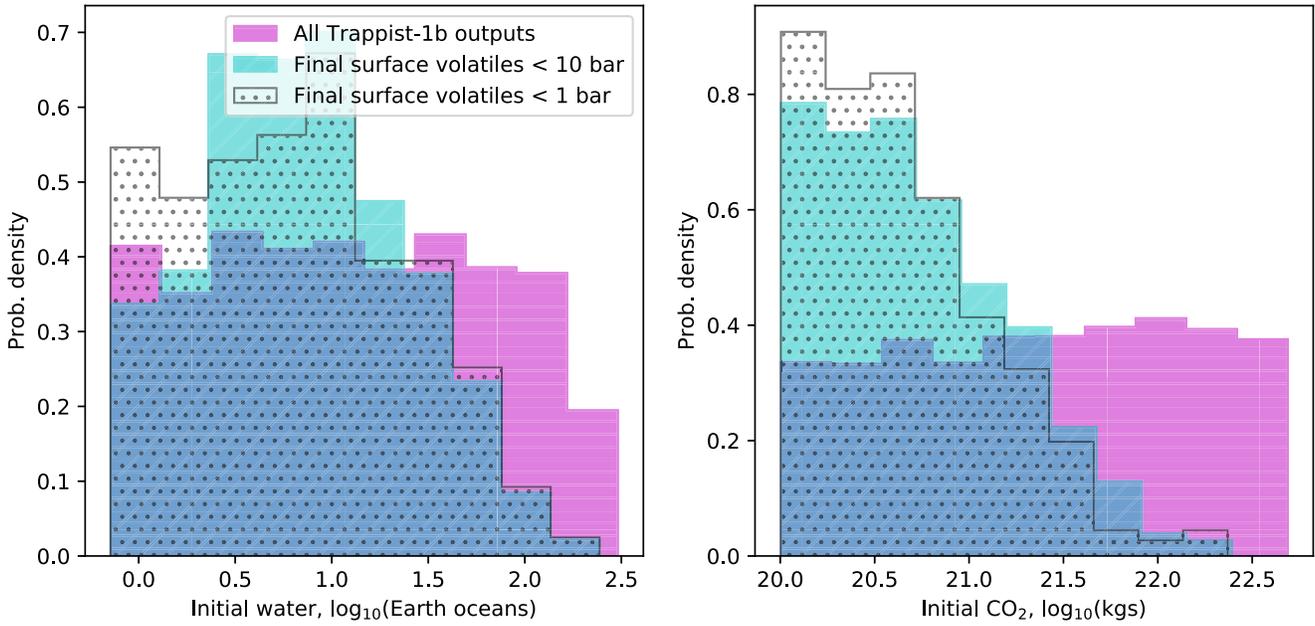

**Figure 4.** Initial volatile inventories for Trappist-1b consistent with the absence of a substantial modern atmosphere. Magenta distributions show initial volatile compositions from all Trappist-1b evolutionary calculations. Cyan and dotted distributions show initial volatile distributions that result in modern surface pressures <10 and <1 bar, respectively. Forward modeling with the PACMAN geochemical model (Figure 1) shows that initial water inventories (left) must have been <∼100 Earth oceans to permit complete atmospheric erosion within the ∼8 Gyr lifetime of the planet. Similarly, initial $CO_2$ inventories (right) must have been less than ∼$10^{21.7}$ kg to permit complete atmospheric erosion. The black dotted (<1 bar) distributions are resampled as the assumed initial volatile inventories for Trappist-1e and f above to represent an endmember case whereby initial volatile endowments within the planetary system are highly correlated.

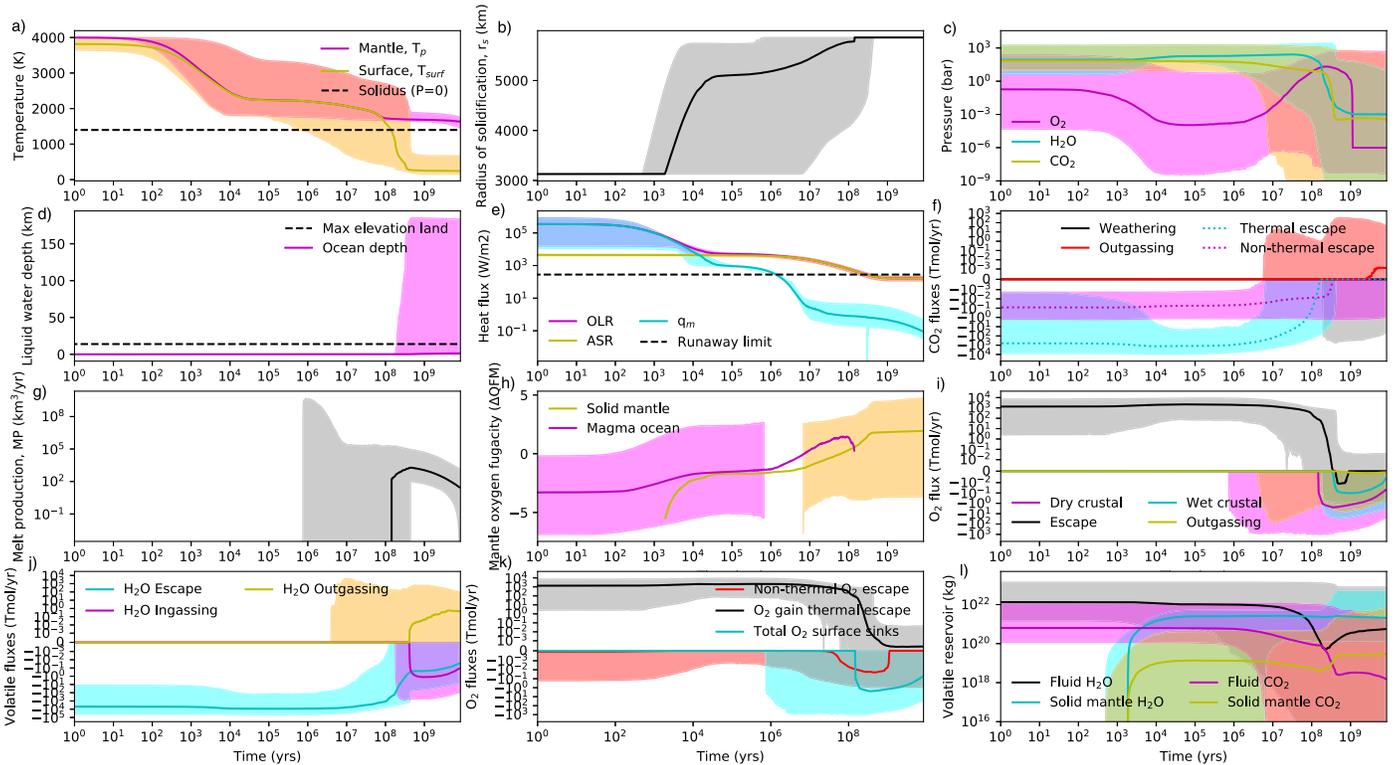

**Figure 5.** All Trappist-1e model outputs consistent with an airless Trappist-1b and c, and with initial volatile endowments constrained by the airlessness of the inner planets. Since Trappist-1e orbits beyond the runaway greenhouse limit during the main-sequence evolution of Trappist-1, the potential for substantial atmospheric erosion is limited, as most model runs in Monte Carlo simulations retain substantial surface volatile inventories.

planet systems, their odds of total atmospheric loss are largely unchanged.

The same holds true even when the initial volatile inventories that produce an airless b are applied to e and f.

This result is attributable to the differing potential for hydrodynamic atmospheric escape on hot planets interior to the runaway greenhouse limit, and habitable zone planets whereby water is expected to be condensed on the surface.

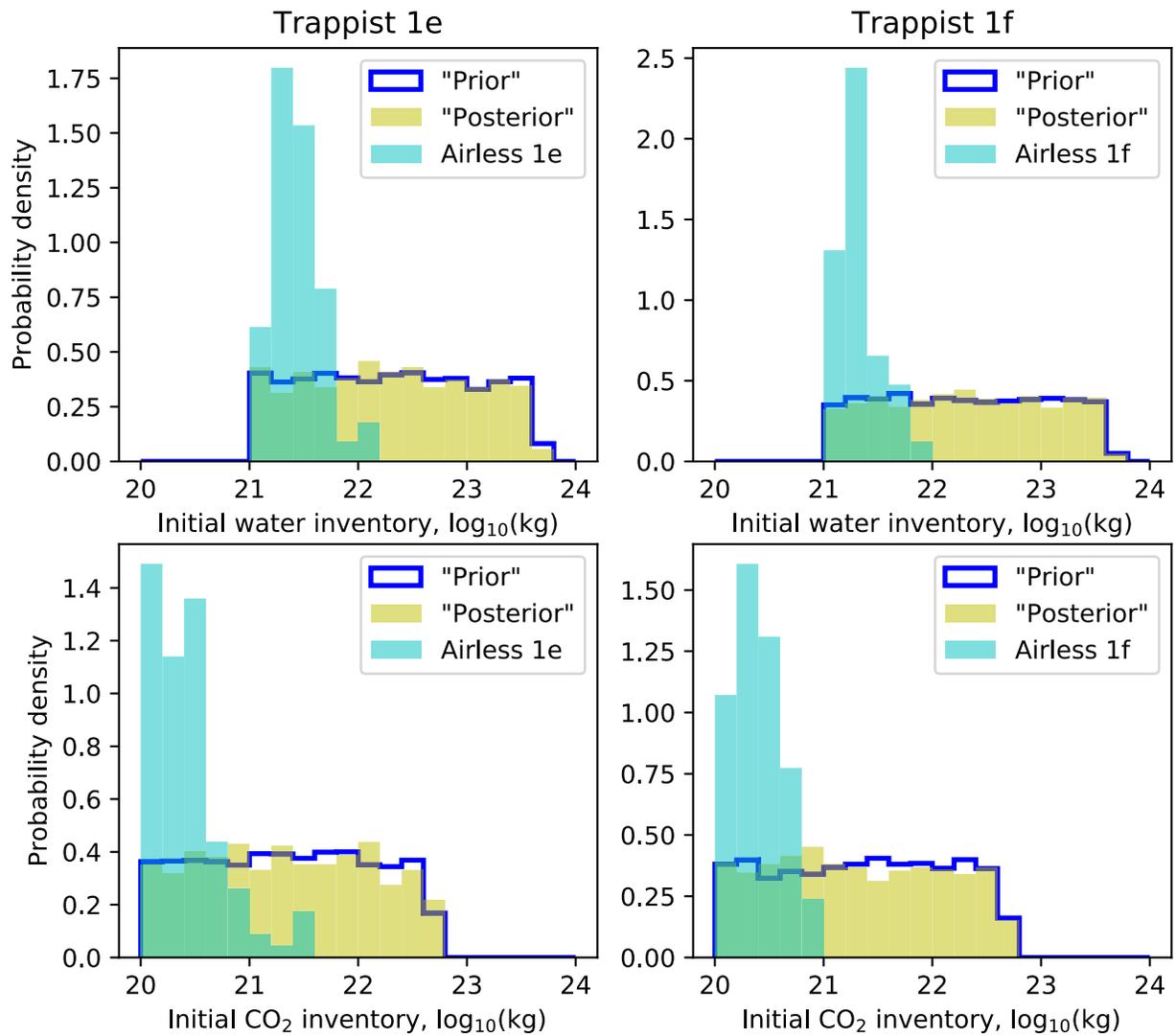

**Figure 6.** Trappist-1e and f initial volatile inventories are unconstrained by an airless 1b and c, but would be tightly constrained if e or f were found to be devoid of surface volatiles. Comparison of initial volatile inventories for Trappist-1e (left) and f (right) assumed Monte Carlo range ("Prior"), implied by an airless Trappist-1b and c ("Posterior"), as well as the distributions required to produce an airless e and f. Top panels denote initial water inventories, whereas bottom panels denote initial $CO_2$ inventories.

While it remains possible that all Trappist-1 planets are airless due to small initial volatile endowments, these results show that the airlessness of b does not require this to be true, and that the anomalously low densities of the outer planets may be attributable to large surface volatile inventories.

### Acknowledgments

I thank Jonathan Fortney, Natasha Batalha, Nicholas Wogan, and the anonymous reviewer for helpful and constructive comments.

### Code Availability

The Python code for our atmosphere evolution model is open source and available at https://github.com/joshuakt/Trappistevolution.